# New Insight Into the Size Tuning of Monodispersed Colloidal Gold Obtained by Citrate Method

*Li Shi,[†,‡] Eric Buhler,[†,‡] François Boué[‡] and Florent Carn[*,†]*

[†] Laboratoire Matière et Systèmes Complexes (MSC), UMR CNRS 7057, Université Paris Diderot, Bâtiment Condorcet, 10 rue Alice Domon et Léonie Duquet, 75013 Paris, France.

[‡] Laboratoire Léon Brillouin, UMR 12 CEA-CNRS, CEA Saclay, 91191 Gif-sur-Yvette, France.

**ABSTRACT.** We study the effect of citrate to gold molar ratio (X) on the size of citrated gold nanoparticles (AuNPs). This dependence is still a matter of debate for $X \geq 3$ where the polydispersity is yet minimized. Indeed, there is no consensus between experiments proposed so far for comparable experimental conditions. Nonetheless, the sole available theoretical prediction has never been validated experimentally in this range of X. We show unambiguously using 3 techniques (UV-Vis spectroscopy, dynamic light scattering and transmission electronic microscopy), 2 different synthetic approaches (Direct, Inverse) and 10 X values for each approach that AuNPs' size decay as a monoexponential with X. This result is, for the first time, in agreement with the sole available theoretical prediction by Kumar et *al.* on the whole studied range of X.

**INTRODUCTION.** Gold nanoparticles (AuNPs) are probably the most widely used and studied metal nanoparticles and have driven a variety of applications in nanomedicine, sensing, optoelectronics and catalysis.[1-5] The control of AuNPs individual characteristics (i.e. size, shape, and size/shape distribution) is fundamental to exploit at a higher length scale their properties that are often related to collective effects.[6] Although, the syntheses in non-aqueous solvents were often preferred for the synthesis of high quality building blocks, great efforts have been done in the last decade to optimize green synthesis pathways directly in water.[7] The proeminent member of this group of aqueous synthesis is probably the 'Turkevich' protocol introduced in 1951 for the synthesis of citrate-stabilized AuNPs.[8] This synthesis enables to obtain quite monodisperse AuNPs in a wide size range by simply changing the relative concentrations of trisodium citrate molecules that are quickly injected in a boiling $HAuCl_4$ aqueous solution.

In contrast with the simplicity of the experimental protocol, the mechanism of AuNPs formation is still obscure on several aspects.[7] This is in part due to the multiple roles played by citrate molecules which result in multiple intricate steps that are hard to probe experimentally. However one can attempt the following basic description. At the beginning, $Au^{III}$ is slowly reduced to $Au^{I}$ thanks to citrate decarboxylation into dicarboxy acetone (DCA) which leads to electron transfer. Ojea-Jiménez and Campanera have shown by differential functional theory simulations that the most favorable reaction path can be decomposed in four steps: (i) substitution of a $Cl^-$ ions by a citrate ligand in the auric acid, (ii) deprotonation of the second most acid carboxylic group, (iii) conversion of the Au equatorial coordination from the initial carboxylate ligand to the hydroxyl group and (iv) formation of transition state.[10] They reveal that the major part of the total activation energy ($\Delta G^{\ddagger}$) corresponds to the two extreme steps and that $\Delta G^{\ddagger}$ decreases from 37.4 kcal/mol to 26.8 kcal/mol when the pH is decreased from neutral to ~ 4. This pH dependence is mainly explained by the fact that $Cl^-$ ions are much more labile around $Au^{III}$ when pH decreases thus facilitating their substitution by citrate.[10]

As formed $Au^I$ atoms may form multimolecular complexes with DCA.[8,11] $Au^0$ atoms could be formed in bulk when the concentration of $Au^I$ species increases to a level 'high enough' ($[AuCl_2]^- \approx 10$ nM [9]) to trigger homogeneous disproportionation.[12] Further disproportionation may leads to formation of still larger aggregates of gold atoms. When the size of the aggregate reaches a critical diameter of about 2 nm, a nucleus of gold atoms may be formed.[13,14] A common feature with other NPs synthesis is that AuNPs with narrow size distribution could be obtained by increasing the speed of this nucleation step. This can be done by favouring the formation of $Au^I$ (i.e. by increasing $[Citrate]_{t=0}/[HAuCl_4]_{t=0} = X$, decreasing pH and/or increasing temperature) and also by increasing the concentration of DCA.[15-20] Once particles are formed, disproportionation can also occurs at particle surface leading to nucleus growth and also to the regeneration of some $Au^{III}$ species.

In contrast with common homogeneous growth of monodisperse particles, several studies *performed at $[Citrate]_{t=0} / [Au]_{t=0} = X \leq 6.7$* have shown that these growing AuNPs could assemble into more or less aniosotropic and crystalline aggregates of several tens of nanometers. [13,15,21-23] The presence of these intermediate aggregates could be understood by considering that several nuclei could be generated by the same $Au^I$/DCA complex. As the reaction proceeds the size of the constituent particles seems to increase until a certain stage at which the aggregates decompose into individual rather monodisperse particles with a typical average diameter of about 15 nm. This mechanism of disaggregation is not readily explained to the best of our knowledge.[9]

Other studies [14,15] show that monodisperse particles can be obtained without such intermediate aggregates. Considering the pH dependence of the standard redox potential of the different $Au^{III}$ complexes and of the reducing agent pointed by Goia and Matijevic,[24] Ji et al. [15] proposed a pH-dependent mechanism of particle formation. It would proceed either in two steps without intermediate aggregates (pH > 6.5), or in three steps, with intermediate aggregates (pH < 6.5).

The final AuNPs are almost spherical when $X \geq 3.5$ in absence of metal contamination.[25] They display an anionic surface charge due to a stabilizing shell composed of deprotonated citrates directly adsorbed on the NP' surface. The dependence of the shell' structure with pH is still matter of discussion. According to a recent study by Park et *al.*[26] the citrate molecules are coordinated by the central carboxylate group at pH = 3.2 which corresponds to $X \approx 1$. Interestingly, they show that this first layer weakly interacting with the gold surface ($E_{COO-/Au} \sim 2$ kcal/mol) could be hydrogen bonded (E ~ 7 kcal/mol for one hydrogen bond of carboxylic acid dimer or 28 kcal/mol for the total citrate interaction) to a second layer of monodeprotonated citrate which give rise to the surface negative charge. The terminal carboxylate groups are progressively coordinated to the gold surface when the pH increases thus suppressing H bond sites and leading to the second layer disappearing.

Following the work of Frens,[27] several recent studies have enable to identify and optimize the main levers (i.e. $[Au^{III}]$,[27,28] X,[8,27-33] [DCA],[15,17,18,20] [electrolyte][34], pH,[15,19,24] temperature[35,36], initiation method[37]) that enable to obtain, eventually in high quantity, AuNPs with radius ranging between 2 and 75 nm with narrow size and shape polydispersity. Among these parameters the reactant ratio, X, is the oldest and probably the most exploited to modify AuNPs size. Forty years ago Frens mentioned a steep decrease of NP' size by a factor 6 when X is varied from 0.8 to 2. This size decrease has been confirmed by several studies[8,27-33] together with the concomitant decrease of the polydispersity. Surprisingly, the experimental results obtained in quite similar conditions are conflicting for $X \geq 3$ (Figure 1) where the polydispersity is yet minimized. Several authors[15,29] has reported a discontinuous size evolution with a sharp minimum at $X \approx 3.5$ while other group[32] indicates that the size decreases continuously with X in an exponential manner on the whole range of X with a lowest radius of ~ 2.5 nm, close to nucleus one, at $X \geq 12$. Beside, the only model that have been proposed so far to rely the AuNP' size to the reactant ratio[33] predicts either a continuous evolution until $R \approx 7.5$ nm at $X = 20$ in absence of AuNP' aggregation or a continuous decrease until $R \approx 10$ nm at $X = 4$ followed by a continuous increase if AuNP' aggregation is taken into account.

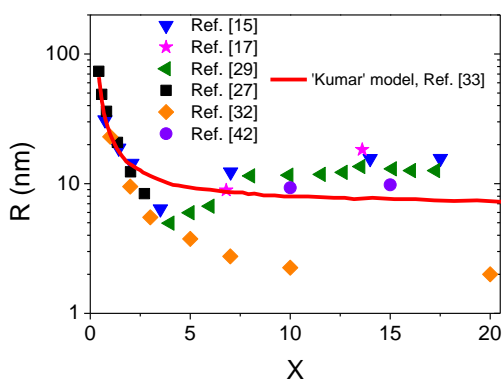

**Figure 1.** Evolution of the averaged radius of citrated AuNPs as function of the molar ratio $X$ = [Citrate]/[Au] for similar conditions of synthesis summarized in table 1 of S.I.. All the radii are number averaged values obtained by TEM except for [29] for who we represent $R_{H, App}$. The line corresponds to Kumar' model prediction.[33]

In this paper, we investigate the relation between AuNP' size and X on a large range of X between 1 and 20 by combining TEM, DLS and UV-Vis. spectroscopy for AuNPs synthesized by the classical Turkevich approach and by the so called 'inverse' method in which the order of reagent injection is reversed.[17,18]

**EXPERIMENTAL SECTION.** Materials. Gold (III) chloride trihydrate (HAuCl$_4$.3H$_2$O, > 99.99 %), trisodium citrate dihydrate (Na$_3$C$_6$H$_5$O$_7$.2H$_2$O, ≥ 99 %) were purchased from Sigma-Aldrich and used as received without further purification. All the content of a gold salt powder batch was used at the first opening to prepare, with glass spatula, a mother solution at 10 g/L in milliQ water that was stocked for period not exceeding 3 months in dark area to minimize photo-induced oxidation. The same batch of trisodium citrate has been used for all synthesis; it has been stored in desiccators after first opening. All glassware and teflon-coated magnetic bars were washed thoroughly with aqua regia and rinsed with milliQ water after each synthesis. All solutions were prepared with milliQ water (R = 18.2 MΩ).

**Chemical synthesis of Citrate stabilized AuNPs.** *"Classical" synthesis.* We took the standard method initially proposed by Turkevich et al.[8]. The gold (III) mother solution was diluted with water to obtain 50 mL of a yellow orange solution at [Au$^{III}$] = 0.25 mM in a 100 mL double-necked round flask. The flask was then immersed in a temperature controlled oil bath at 96 ± 3 °C without reflux and was vigorously stirred at ~900 rpm. 10 min after reaching the desired temperature, 2.5 mL of a citrate solution (final X ranging between 1 and 20) preheated during 10 min at the reaction temperature were all added at once with a thermalized tip. After 15 min under vigorous stirring at 96 °C, we switch off the stirring and the oil bath. The magnetic bar is removed but not the oil bath until reaching ∼ 70 °C (i.e. after 20 min). Then, the solution was allowed to cool down to the room temperature. The time evolution of the oil bath' and solution' temperatures measured during a 'classical' synthesis is plotted in the figure 3 of S.I. Visually, the initially yellow orange colored solution containing Au$^{III}$ species turned clear, as a result of the partial reduction to Au$^I$ complexes that did not adsorb in the visible range,[24] over dark blue, due to the formation of Au$^0$ atoms, and finally left a deep wine red color within several minutes indicating the formation of 'nanometric' AuNPs.

*"Inverse" synthesis.* We used the protocol proposed by Ojea-Jimenez et al.[17]. Desired amount of Citrate were diluted in a 250 mL double-necked round flask with water and stirred during 20 min at 25 °C. The solution was then heated during 15 min at 100 °C ($T_{oil\ bath}$ = 140 ± 5 °C) with reflux and vigorous stirring at 900 rpm. Then, one milliliter of mother precursor solution ([Au] = 25 mM) preheated at 100 °C during 10 min was injected all at once with a thermalized tip into the citrate solution under the same vigorous stirring. After 15 min at 100 °C, we switch off the stirring and the oil bath. The magnetic bar is removed but not the oil bath until the solution's temperature had reached ~ 70 °C (i.e. after ~ 40 min). Then, the solution was allowed to cool down to the room temperature. We show in Figure 1 of S.I. the evolution of pH and electrical conductivity for AuNPs prepared by the "Inverse" method with X ranging between 1 and 20. Each synthesis has been reproduced 3 times with the same protocol and set-up.

**Methods.** *Dynamic light scattering (DLS)* experiments were carried out using on a NanoZS apparatus (Malvern) operating at λ = 632.8 nm. The dependence of the characteristic relaxation times with the scattering angle (θ) was determined with a 3D DLS set-up (LS Instruments) operating at λ = 632.8 nm. The dispersions were never filtered before measurements.

For dispersions characterized by two diffusive relaxation mechanisms, the electric field autocorrelation function $g^{(1)}$(q, t) can be fitted with the sum of two exponentials:

$$g^{(1)}(q,t) = A_{fast}(q)e^{-t/\tau_{fast}} + A_{slow}(q)e^{-t/\tau_{slow}}$$

where $\tau_{fast}$ and $\tau_{slow}$ represent the two cooperative characteristic relaxation times, and $A_{fast}$ and $A_{slow}$ are their corresponding amplitudes with $A_{fast}(q) + A_{slow}(q) = 1$.

*Electrophoretic mobility (µ)* was also measured with a NanoZS (Malvern Instrument). This set-up operates with an electrical field of 25 V.cm$^{-1}$ oscillating successively at 20 Hz and 0.7 Hz to reduce the electroosmosis effect due to the surface charge of the capillary cell. The particle' velocity is measured by LASER Doppler velocimetry.

*Ultraviolet visible spectroscopy (UV-vis)* was performed on Cary 50 Scan UV-Visible Spectrophotometer of brand Varian. The diluted AuNPs were filled in the 5mm thickness Hellma cell (quartz). The absorbance values were recorded after baseline correction.

*Transmission electron microscopy (TEM)* experiments were performed on a JEOL 2010 instrument operating at 200 kV. Samples were prepared by casting a single drop of a 1g/L aqueous dispersion onto a standard carbon-coated Formvar films on copper grids (200 mesh) and drying in air for more than 30 min. A minimum of 250 NPs per sample were considered for the statistics except at X = 1 for which 50 NPs were considered.

**MODEL.** In this section we briefly present the main elements of the model proposed by Kumar et al.[33]. To our knowledge, this is the only model proposed so far to describe the formation of citrated AuNPs.

*Main hypothesis.* The model considered that: (i) the critical nucleus' size is about 2 nm, (ii) all the AuNPs including those formed by aggregation are considered to be spherical, (iii) the repulsive interaction between AuNPs is only due to double layer interaction, (iv) the rate of AuNPs' aggregation could be derived from the expression introduced by Reerink and Overbeek[38] (v) AuNPs' aggregation is irreversible.

This model does not take into account: (vi) the possible disaggregation of intermediate aggregates' leading to individual AuNPs, (vii) the hydrothermal oxidation of citrate and (viii) the pH evolution as a function of the molar ratio X and the inherent effect on the electrochemical potential of the dominant aureate complexes in water, (ix) the steric repulsion due to citrate adsorption on the AuNPs surface but the author balances this by overestimating the AuNPs' surface potential on purpose. The value of the later is hence determined by an 'empirical' relation giving values always higher than the 90 mV derived from electrophoretic mobility measurements performed at X = 3 by Chow and Zukoski.[13] We show in figure 2 of the S.I. that our AuNPs have almost the same electrophoretic mobility than the later in the same range of X.

*Scheme of reactions.* The model considers that the formation of citrated AuNPs result from the following reactions involving auric and aurous species (noted T and M respectively); citrate, dicarboxy acetone and acetone (noted C, DCA and A respectively).

(1) Homogeneous reduction auric ions by citrate: $T + C \xrightarrow{k_1} M + DCA + products$

(2) 'Degradation' of dicarboxyacetone: $H_2O + DCA \xrightarrow{k_2} A + products$

   This reaction corresponds to the hydrothermal oxidation of dicarboxy acetone.

(3) Homogeneous reduction by acetone: $4\,T + A \xrightarrow{k_3} 4\,M + products$

   Note that the stoichiometric ratio of this reaction have been assigned to obtain a complete conversion of auric chloride at a stoichiometric ratio of 0.4 as observed by Frens.[27]

(4) Nucleation: $3\,M + 2\,DCA \xrightarrow{k_4} T + Nucleus + 2\,DCA$

(5) Heterogeneous disproportionation: $3\,M \xrightarrow{k_5} T + particle\ mass$

   The rate of this reaction is assumed to be proportional to the surface area of the AuNP since the gold surface catalyzes this reaction. The number of gold particles in a given size range R to R + dR hence appears in the mass balance of aurous and auric species.

The number of particles in a size range as a function of time is obtained by solving the 'population balance equation' using a discretization technique described by the same author in [39]. The 'population balance equation' is obtained by considering that the apparition and disapearance of AuNPs in a given interval of sizes results from growth process due to surface reaction (3) and Brownian aggregation while new nuclei are formed in continuous (reactions (1), (3) and (4)) until the concentration in DCA is insufficient due to the degradation process (2). The rate constant $k_1$ was fixed to make the process time of the same order as the experimentally reported values at 100 °C while the other rate constants were adjusted to obtain the best fitting of the AuNP' sizes, determined with TEM by [8,27,30,31] variation with X in the range: $0.4 \leq X \leq 7.5$. The initial gold concentration is fixed at $[T]_{t=0}$ = 0.3 mM and the initial citrate concentration $[C]_{t=0}$ is varied to cover an X range between 0.4 and 40.

**RESULTS AND DISCUSSION.** AuNPs' dispersions prepared by the Turkevich (i.e. "classical") and "inverse" methods were first characterized in bulk by UV-Vis. spectroscopy as shown in figure 2. Whatever the method of preparation, one observes a significant narrowing of the surface plasmon resonance (i.e. SPR) peak accompanied by a blue shift of SPR maximum from 546 nm (direct synthesis) / 555 nm (inverse synthesis) down to 519-520 nm when X increases from 1 to 3.5 (figure 2c). Then, the SPR peak does not evolve up to X = 8. Above this value a slight red shift can be detected up to 522-523 nm at X = 20. This evolution agrees qualitatively with the picture of a transition from large and polydisperse AuNPs at X < 3.5, to small and quite monodisperse AuNPs at X ≥ 3.5 until a limit at high X values above which aggregation may occurs. We tried to derive the intensity average AuNPs' size from these absorbance data using the following relation proposed by Haiss et al.[40]: $<R>_{Int.} = \frac{1}{2} e^{\left(3 \cdot \frac{A_{SPR}}{A_{450}} - 2.2\right)}$ where $A_{SPR}$ and $A_{450}$ are the absorbance at the SPR peak and 450 nm respectively, while the factors 3 and 2.2 were empirically determined by the authors. The as-determined radii evolution plotted in figure 2.c give rise to a rather unaccurate result due to points noticeably scattered in the whole range of X. It appears that the Haiss' relation is too much sensitive to uncorrelated variations of the absorbance and does not enable to extract in a quantitative manner the evolution of the AuNPs' sizes.

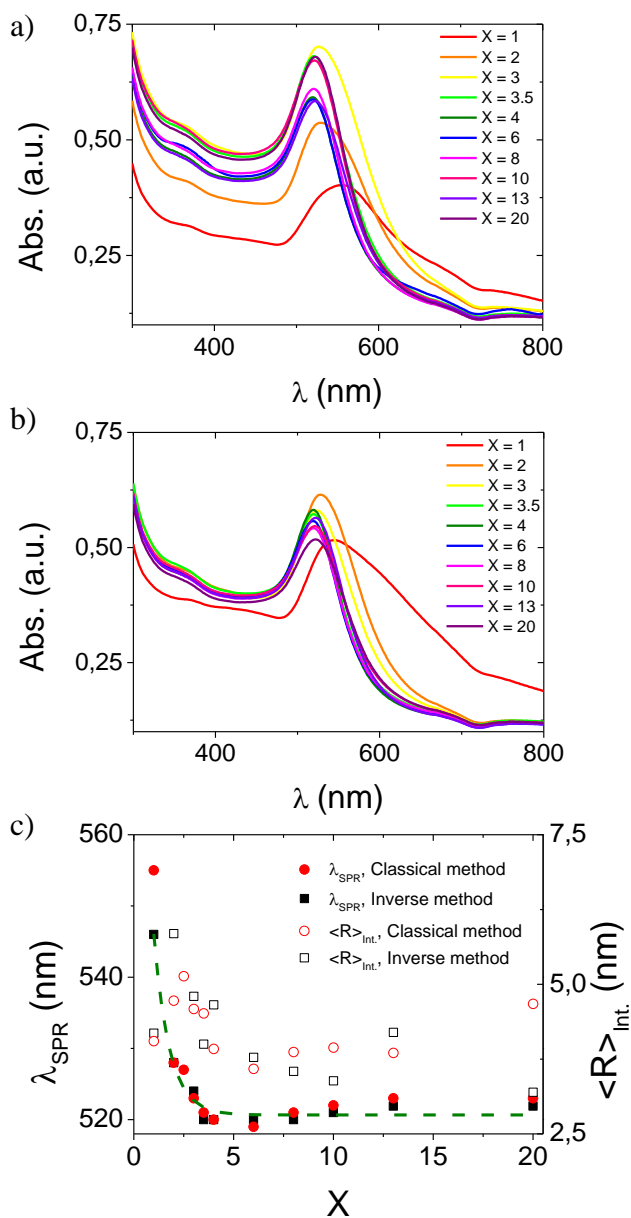

**Figure 2.** Absorbance (Abs.) vs. wave length ($\lambda$) at 25 °C for AuNPs dispersions synthesized by (a) "Classical" and (b) "Inverse" methods for different ratio X. (c) Evolution of the wave length at the surface plasmon resonance ($\lambda_{SPR}$, full symbol) and AuNPs' intensity averaged radius

(<R>$_{Int.}$, open symbol), determined with Haiss relation,[40] for "Classical" (red circle) and "Inverse" (black square) methods. The dashed line is an exponential decay (i.e. $\lambda_{SPR}(X) = 90 \times e^{-X/0.79} + 520.6$) and must serve as guide for the eyes.

To better quantify the size distributions in bulk we performed DLS measurements. The normalized electric field autocorrelation functions, $g^{(1)}$ (173°, τ), measured one day after the synthesis are plotted in figure 3. For the two synthetic routes, the curves depict two relaxations for X = 1 and one relaxation for: 1 < X < 10 with the classical method or 1 < X < 20 with the 'inverse' method. In these intermediate ranges, the relaxation times decrease when X goes from 2 to 3.5 and no longer varies beyond. At high X ratio, these relaxations are stretched or associated with a slow mode of low amplitude (i.e. X = 13 or X = 20 for classical and inverse methods respectively). This analysis suggests again the formation of particles of smaller and smaller size until X = 3 after what the size does not evolve significantly except at high X values where aggregation may appears.

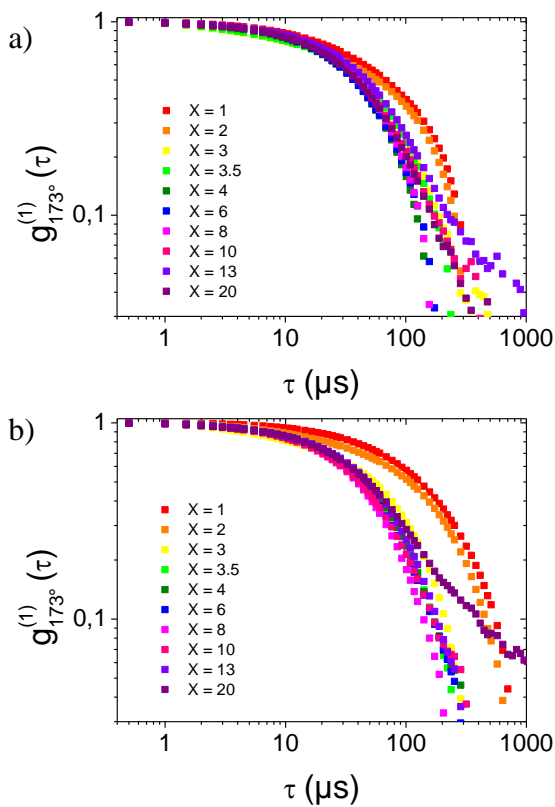

**Figure 3.** Normalized electric field autocorrelation functions, $g^{(1)}$ (173°, τ), obtained at 25 °C, one day after (a) classical or (b) inverse synthesis with different ratio X indicated on the plot.

Accordingly, the intensity weighted distributions of apparent hydrodynamic radius ($R_{H, App.}$) obtained with the Contin' algorithm (Figure 4) reveal a transition from large and polydisperse AuNPs at X ≤ 3, to small and quite monodisperse AuNPs at X ≥ 3.5, with:

$<R_{H, app.}>_{3.5 \leq X \leq 20}$ = 9.9 ± 1.2 nm for the classical method and $<R_{H, app.}>_{3.5 \leq X \leq 20}$ = 10.6 ± 1.5 nm for the 'inverse' method.

Interestingly, at high X values, the presence of a second population of large objects with $R_{H, app}$ > 100 nm is not associated with a modification of the small objects' mean size. In spite of their big size (hence large individual scattering) the large objects detected at X = 13 and X = 20 for classical and 'inverse' methods respectively are always minority in % of intensity meanwhile they represent a low number fraction compared to the one of individual NPs. This can be quantified by considering the amplitudes $A_{fast}$ and $A_{slow}$ associated respectively to the fast and the slow mode (see E.S.). Indeed, defining $I_{indiv.}$ and $I_{aggregates}$ as the time-averaged intensities associated with the fluctuations of individual and aggregated AuNPs concentrations respectively, and neglecting the virial effects, we obtain:

$$\frac{A_{fast}}{A_{slow}} \sim \frac{I_{indiv.}}{I_{aggregates}} \sim \frac{K.C_{indiv.}M_{w,indiv.}}{K.C_{aggregates}M_{w,aggregates}} \sim \frac{C_{indiv.}}{C_{aggregates}} \left(\frac{R_{H,indiv.}}{R_{H,aggregates}}\right)^3$$

Where K is the scattering constant, $M_{w,indiv.}$ and $M_{w,aggregates}$ are respectively the molecular mass, in g.mole$^{-1}$, of individual and aggregated AuNPs while $C_{indiv.}$ and $C_{aggregates}$ are respectively the concentrations, in g.cm$^{-3}$, of individual and aggregated AuNPs.

Hence, we obtain: $\frac{C_{indiv.}}{C_{aggregates}} \sim \frac{A_{fast}}{A_{slow}}\left(\frac{R_{H,aggregates}}{R_{H,indiv.}}\right)^3 \sim 5.10^4$ for X = 13 in the 'classical' method and $\frac{C_{indiv.}}{C_{aggregates}} \sim 5.10^5$ for X = 20 in the 'inverse' method.

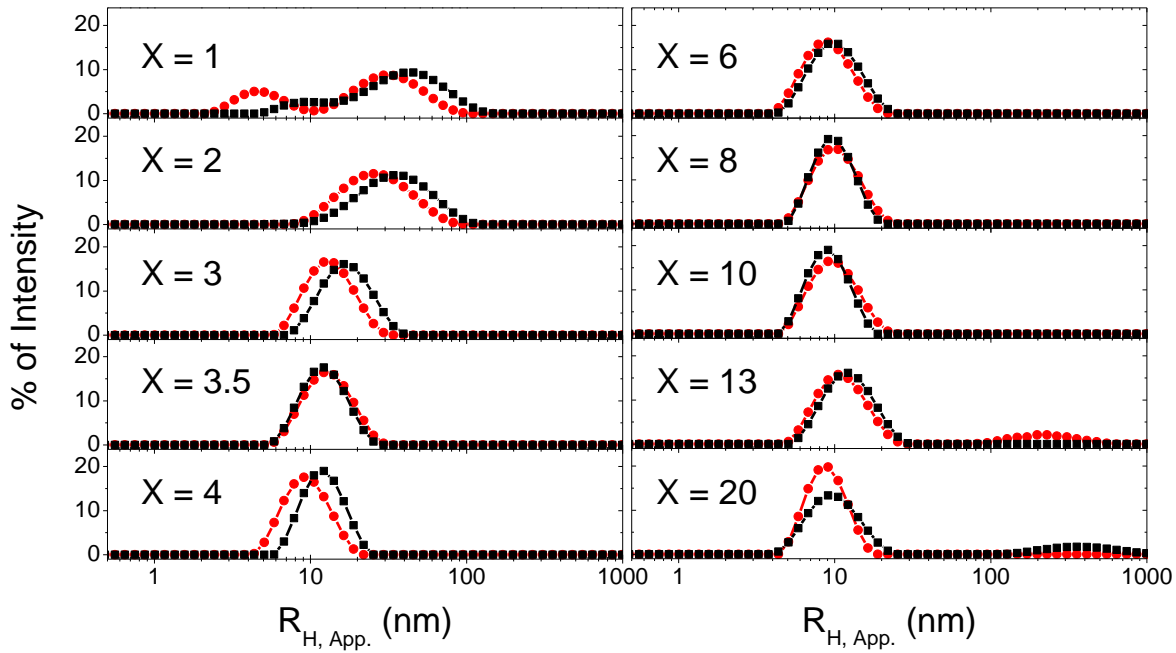

**Figure 4.** Intensity weighted distributions of $R_{H, App.}$ for AuNPs prepared by classical (red disc) and inverse methods (black square) at different ratio X indicated on the plot.

The number weighted size distributions have been then determined for the two sets of synthesis by means of TEM. Overall, these distributions present the same evolution of size and polydispersity with X than the distribution of $R_{H,\ app.}$ except the link between the presence of aggregates and the high X values. that cannot be revealed by this method. Indeed TEM implies here sample drying which may disperse weak aggregates as well as create aggregates (from dewetting for example). However, it is possible to follow the shape of the individual NP among the apparent aggregates (or individual NPs). We see the progressive shape transition from anisotropic (X = 1, X =2) to isotropic AuNPs at X > 3 already reported by several authors for a threshold 3.5. The number averaged size of the isotropic NPs obtained at X ≥ 3.5 is: $<R_{TEM}>_{3.5 \leq X \leq 20}$ = 8.2 ± 1.0 nm for the classical method and $<R_{TEM}>_{3.5 \leq X \leq 20}$ = 8.3 ± 1.3 nm for the 'Inverse' method.

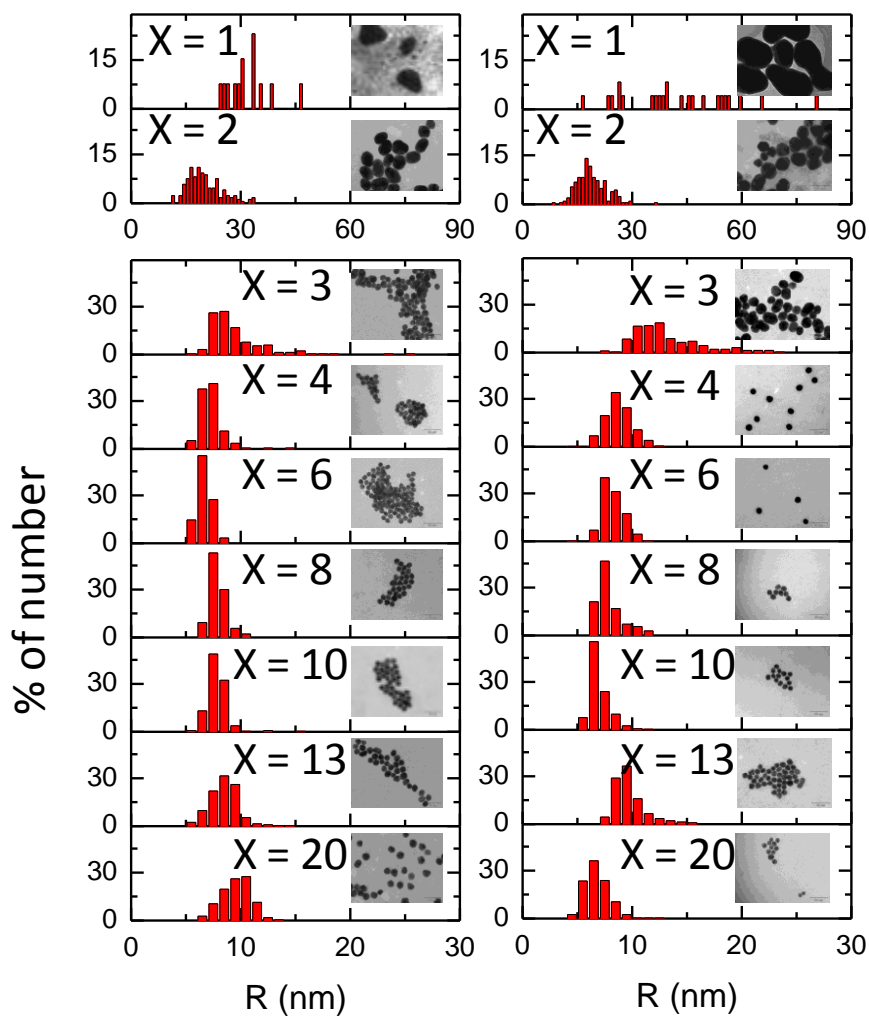

**Figure 5.** Number weighted size distribution determined by TEM observations of AuNPs prepared by the traditional method (left column) and the inverse (right column) at different X ratio indicated on the plot. Representative TEM images taken with the same magnification are shown as inset for each histogram. The width of the images corresponds to 560 nm.

Figure 5 summarizes all the results obtained during this study. It appears that the intensity average radii determined by DLS and TEM decrease in the same manner when X increases for the two batch of AuNPs. We consider it as a good support for reliability of the approach. The size evolution is sharp at low X and not significant above X = 3.5. This prompts us to a more detailed focus on the slight increase of $\lambda_{SPR}$ (red stars in the plot of Figure 6) for X ≥ 3.5, and to consider that it should not be assigned to an increase of AuNP' size.

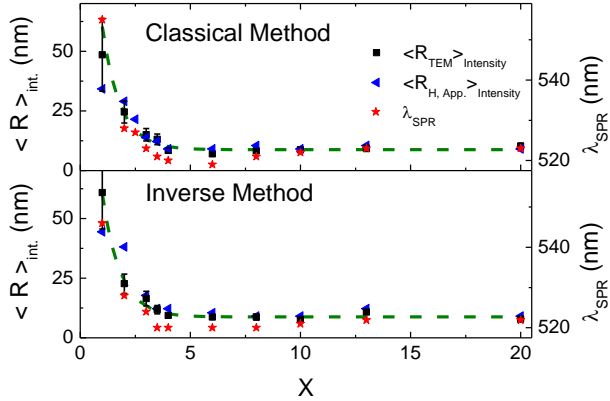

**Figure 6.** Evolution of the $\lambda_{SPR}$ and the intensity average radii determined by DLS and TEM, $<R>_{int.}$, as function of X for the two methods of synthesis as indicated in the figure. For the two plots, the dashed line is the same exponential decay function (i.e. $\langle R(X) \rangle_{Int.} = 176 \times e^{-X/0.8} + 8.8$ ) and must serve as guide for the eyes.

The comparison between our measurements and previous, experimental[15,29,32] and theoretical[33], data is shown in figure 7. One observes a good agreement between our results and the theoretical prediction derived from Kumar' model in contrast with all other experimental results available so far for comparable experimental conditions, to the best of our knowledge. Looking closer, it could be mentioned that the data corresponding to the 'inverse' method of synthesis are peculiarly well fitted by the model that do not consider particle aggregation except at low X values where our data are slightly shifted. This may be related to the hydrothermal oxidation of citrate preformed before $Au^{III}$ addition which is not considered in the Kumar' model designed for the classical method where this process may be neglected in the presence of the strongly oxidant $Au^{III}$ species. Beside this effect, the semi-log representation highlights that size dependence for AuNPs prepared by the classical method is surprisingly less well captured by the model than for the 'inverse' method.

This result leads us to underline the qualitative character of the nice agreement between our data and the Kumar' prediction considering the number of basal hypothesis of this model and the fitting procedure based on an adjustment of rate constants that remained to be measured.

However, in light of this agreement one can recall that Kumar model attribute the AuNPs' size dependence with X to a balance between the rates of degradation of DCA and of nucleation. The sharp decrease of size observed at low X should be due to a sharp increase in the number of nuclei formed as the initial concentration of citrate is increased for a fixed concentration of gold. This effect may be explained by the presence of a maximum in the time evolution of DCA' concentration resulting from the competition between the mechanisms that trigger DCA formation / degradation. The rate of nucleation should be maximal at this moment, due to the implication of DCA in the nucleation process. Therefore, increasing the initial citrate concentration ($[C]_{t=0}$) at fixed gold concentration, should increase the maximum concentration of DCA and hence the nucleation rate. This process should occur until a limit at which auric chloride became the limiting reactant of the nucleation process. Beyond this limit, which is experimentally observed for us around X = 3.5, the nucleation rate became almost independent of $[C]_{t=0}$. In this range of X, Kumar' model predicts that AuNPs' size should be relatively independent of X without aggregation, or instead, should increase at high X (i.e. X ≥ 8) due to aggregation induced by the increase of ionic strength that get along with the increase of X. Interestingly, our results slightly differ from these two options since we observe that AuNPs grow to their final size without significant effect of aggregation for X ≥ 3.5 (i.e. no significant dependence of <R> with X) and could be aggregated at a much higher length scale ($R_{H, app.}$ >100 nm) for high X values. Note that these aggregates could result from the aggregation of preformed individual AuNPs as considered by the model or from an incomplete disaggregation of initially interconnected AuNPs.

We emphasize that the effect of citrate on pH and the correlated effect on the electrochemical potential of the dominant aureate complexes in water is not considered by Kumar et *al*. Nonetheless recent experimental and theoretical studies has pointed that the reduction was facilitated for [AuCl$_4$]$^-$ and highly deprotonated citrate ions with the gold species having the major impact on the total reaction free energy. Considering that gold equilibrium is shifted from [AuCl$_4$]$^-$ to less reactive [AuCl$_{3-x}$(OH)$_{1+x}$]$^-$ during the fast seed particle formation (~ 30 s after mixing), Wuithschick et al.[41] propose that if citrate is added in a sufficient concentration, the kinetics of the protonation' equilibrium and the final pH of the solution are approximately the same irrespective of the molar excess thus explaining the size independence with X above X = 3.5. Below this limit, the molar excess might still be sufficient to reduce the amount of Au$^{3+}$ but not to shift of the gold complex equilibrium from reactive [AuCl$_4$]$^-$ to less reactive hydrolyzed forms. This agrees with the observed pH evolution with X (Figure 1 of S.I.). In this condition beside the reduction of Au$^{3+}$ in the electronic double layer of preformed particles, reduction can also occur unselectively during the entire synthesis leading to the nonuniform polydisperse final colloids observe at low X value.

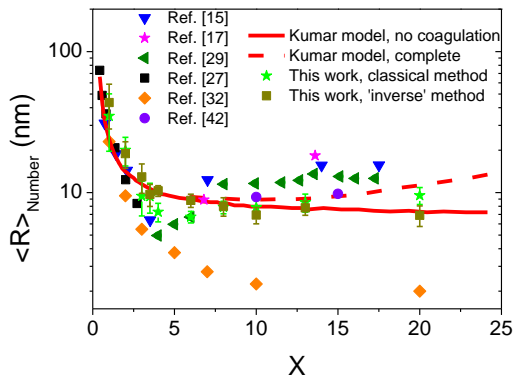

**Figure 7.** Evolution of the average radii of AuNPs prepared in similar conditions of synthesis summarized in table 1 of S.I., as function of the molar ratio X. All the radii are number averaged values obtained by TEM except for [29] for who we represent $R_{H, App.}$. The lines correspond to: Kumar model prediction with (dashed line) and without (continuous line) aggregation.[33]

Thereby, in contrast with previous experimental studies,[15,29,32] our measurements agree with Kumar' prediction on: (i) the absence of sharp and deep minimum of size at X ≈ 3.5 and (ii) the absence of continuous size decrease until a size close to nucleus one at X ≈ 20. This dispersion of experimental results for AuNPs prepared in conditions that are *a priori* comparables (i.e. T ≈ 100 °C, variation of X for $[Au^{III}]_{t = 0}$ ≈ 0.25 mM : fixed) is puzzling. However, we notice that several experimental details regarding the synthesis (i.e. the use of reflux, the volumes of each reactant, the bath temperature, and the time of mixing at each step) and the characterizations (i.e. number of NPs considered in the TEM statistics, plot of $g^{(1)}$ (q, τ) determined by DLS) are often missing in the literature thus hindering a detailed analysis of this results' dispersion.

Concerning the effect of the citrate addition order, our results do not reveal a significant gain in size monodispersity when 'inverse' method is preferred to classical one for X ≥ 4.

**CONCLUSIONS.**

In conclusion, we report that the size of AuNPs prepared by 'classical' Turkevich-Frens approach and, also, by the 'inverse' approach, proposed by Ojea-Jimenez *et al.* and by Sivaraman *et al.*, decay in a mono exponential manner when the molar ratio X = [Citrate]$_{t=0}$ / [Au$^{III}$]$_{t=0}$ increase between 1 and 20. This result has been established by an unambiguous set of experimental results (UV-Vis spectroscopy, dynamic light scattering and transmission electronic microscopy). In contrast with all other previous experimental studies, we show that the reported results are in good agreement with the theoretical prediction by Kumar et *al.* in absence of AuNPs aggregation on the whole range of X.

The dispersion of experimental results for AuNPs prepared in conditions that are *a priori* comparables is puzzling. We point in the article that important informations concerning the synthesis and the characterizations are unfortunately often missing in the literature thus hindering a detailed analysis of the results' dispersion.

ASSOCIATED CONTENT

**Supporting Information**. Table summarizing the experimental conditions of the synthesis proposed in [8, 15, 27-29, 32]. Evolution of pH, electrical conductivity and electrophoretic mobility as a function of X. Typical, low magnification TEM images of AuNPs prepared by "Classical" and "Inverse" method for X ranging between 1 and 20. "This material is available free of charge via the Internet at http://pubs.acs.org."

AUTHOR INFORMATION

**Corresponding Author.**

**\* E-mail:** florent.carn@univ-paris-diderot.fr

ACKNOWLEDGMENTS. L.S., E.B. and F.C. are indebted to the French Labex SEAM (Sciences and Engineering for Advanced Materials and devices) supported by the 'Commissariat Général à l'Investissement'.

**Table of Contents Graphic.**

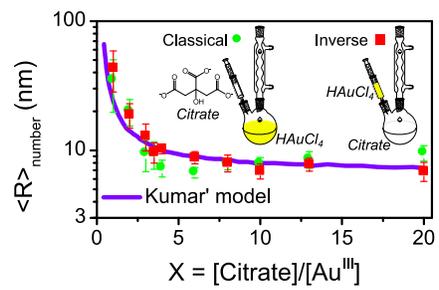